\documentclass[12pt]{article}
\pdfoutput=1
\usepackage[utf8]{inputenc}
\usepackage[english]{babel}
\usepackage{amssymb,graphicx}
\usepackage{amsmath,amsfonts}
\usepackage{fullpage}

\newcommand{\be}{\begin{equation}}
\newcommand{\ee}{\end{equation}}
\newcommand{\bea}{\begin{eqnarray}}
\newcommand{\eea}{\end{eqnarray}}

\title{
\vspace{-2cm} 
\begin{flushright}
{\normalsize INR-TH-2021-014}
\end{flushright}
\vspace{0.5cm} 
Two-sided constraints on Lorentz Invariance violation from Tibet-AS$\gamma$ and LHAASO Very-High-Energy photon observations}
\author{Petr Satunin\thanks{{\bf e-mail}: satunin@ms2.inr.ac.ru}
\vspace{.2cm}\\
\normalsize\it  Institute for Nuclear Research of the Russian Academy of Sciences, \\  
\normalsize \it  60th October Anniversary Prospect, 7a, 117312  Moscow, Russia} 
\date{}
\begin{document}
\maketitle
\begin{abstract}
We present new two-sided constraints on the Lorentz Invariance violation energy scale for photons with  quartic dispersion relation from recent gamma ray observations by the Tibet-AS$\gamma$ and LHAASO experiments. The constraints are based on the consideration of the processes of photon triple splitting (superluminal scenario) and the suppression of shower formation (subluminal). The constraints in the subluminal scenario are better than the pair production constraints and are the strongest in the literature. 
\end{abstract}

\paragraph{Introduction} 
Since 2019, a new window in Very-High-Energy (VHE) astronomy focusing on gamma-rays with energy from $100$ TeV to several PeV has been opened. The photon spectra of galactic sources which continue beyond  $100$ TeV  have been measured by the collaborations Tibet-AS$\gamma$ \cite{Amenomori:2019rjd}, HAWC \cite{Abeysekara:2019edl, Abeysekara:2019gov, Albert:2020yty, Abeysekara:2021yum}, LHAASO \cite{Aharonian:2020iou, LHAASO} and Carpet-3  \cite{Dzhappuev:2021cfj}. In addition, the Tibet-AS$\gamma$ collaboration has reported  the detection of a diffuse photon flux from the  Galactic disk \cite{Amenomori:2021gmk}.  Two of these collaborations, Tibet-AS$\gamma$ \cite{Amenomori:2021gmk} and LHAASO \cite{LHAASO}, have reported  single photon events with energy near PeV --- $957$ TeV and $1.42$ PeV, respectively. It is expected that these recent experimental data will shed a lot of light on the mechanism of cosmic rays acceleration in galactic sources. 
In addition, the physics of these gamma rays may be sensitive to modification of some fundamental laws of physics, such as violation of Lorentz Invariance (LI). 

The motivation for considering broken LI was born in several different attempts to construct quantum gravity theory (see \cite{Mattingly:2005re,Liberati:2013xla} and references therein for a review). 
Lorentz Invariance Violation (LIV) in the matter sector is usually phenomenologically considered 
in the framework of Standard Model Extension \cite{Colladay:1998fq} (SME). The main feature of LIV is the modification of dispersion relation for particles \cite{Coleman:1997xq}. This results in kinematical effects such as  the modification of thresholds for some processes in particle physics \cite{Coleman:1997xq,Jacobson:2002hd}.  However, the dynamical effects of LIV including the modification of cross-sections for non-threshold processes are also important. The processes important for LIV photon propagation and detection in the atmosphere include photon decay \cite{Coleman:1997xq,Jacobson:2002hd,Rubtsov:2012kb}, photon triple splitting \cite{Gelmini:2005gy,Astapov:2019xmt}, pair production on soft photon background \cite{Kifune:1999ex, Stecker:2001vb} and in the Coulomb field of nuclei \cite{Vankov:2002gt, Rubtsov:2012kb}. The last two processes hold in the standard case and modified in a LIV scenario. The modern constraints on LIV in the photon sector are gathered in the Data Tables \cite{Kostelecky:2008ts}, see also the recent review \cite{Martinez-Huerta:2020cut}.

On this date, several constraints on LIV parameters in the photon sector based on LHAASO data have been already obtained in \cite{Li:2021tcw} and \cite{Chen:2021hen}\footnote{After the first version of the current paper the LHAASO collaboration article on constraining LIV appeared \cite{LHAASO:2021opi}.}. The article \cite{Li:2021tcw} is devoted to constraining cubic corrections to the photon dispersion relation while the article \cite{Chen:2021hen} is devoted to quartic correction, focusing on the processes of photon decay and photon splitting. In this Letter we complement these works with Tibet-AS$\gamma$ data as well as with the constraints from air shower formation.



\paragraph{The Model.}
In this Letter we focus on quartic corrections to the photon dispersion relation, 
\begin{equation}
\label{DispRelation}
    E_\gamma^2 = p_\gamma^2  \pm \frac{p_\gamma^{4}}{M_{LV}^2}.
\end{equation}
Here $M_{LV}$ denotes the energy scale of LIV quantum gravity\footnote{In the  literature the notation $E_{QG,(2)}$ is also used.}.
The sign plus (minus) in the dispersion relation (\ref{DispRelation}) refers to a superluminal (subluminal) LIV scenario.
The dispersion relation (\ref{DispRelation}) can be obtained from the Lagrangian,
\begin{equation}
\label{L1}
\mathcal{L}=-\frac{1}{4}F_{\mu\nu}F^{\mu\nu}  \mp \frac{1}{2 M_{LV}^2}F_{ij}\Delta^2 F^{ij}  +i\bar{\psi}\gamma^\mu D_\mu\psi - m\bar{\psi}\psi.
\end{equation}
Note that Feynman rules for photons in the model (\ref{L1}) are modified, see \cite{Rubtsov:2012kb} for details.

\paragraph{Photon decay.} The simplest process characteristic to a superluminal LIV scenario, is the photon decay to an electron-positron pair $\gamma \to e^+ e^-$. The process has a certain energy threshold: below the threshold it is forbidden by the energy-momentum conservation; for energies larger than the threshold one the decay rate is very fast\footnote{The mean free path for TeV photons is about one hundred meters.} \cite{Rubtsov:2012kb}. 
The detection of even a single photon of a given energy means that the energy is larger than the threshold one and so establishes the constraint,
\begin{equation}
\label{PhDecay}
M_{LV} > \frac{E_\gamma^2}{2m_e}.    
\end{equation}
The significance of the constraint (\ref{PhDecay}) coincides with the significance of the corresponding photon event.

\paragraph{Photon splitting.} In the superluminal scenario $\gamma \to e^+e^-$ is not the only channel of the photon decay. The process of photon splitting to three photons $\gamma \to 3\gamma$ occurs without threshold if the photon dispersion relation is superluminal, so if the photon decay is kinematically forbidden the splitting may be the most effective process. The splitting width reads \cite{Astapov:2019xmt},
\begin{equation}
\label{FinalGamma}
\Gamma_{\gamma\to 3\gamma}\ \simeq\ 5\cdot 10^{-14}\;\frac{E_{\gamma}^{19}}{m_e^8M^{10}_{LV}}.
\end{equation}
The splitting process starts making a visible effect when the photon mean free path associated with splitting is of the same order as the distance to the source, $\left( \Gamma_{\gamma\to 3\gamma}\right)^{-1}\,\sim L_{source}$. 
The probability for a given photon to pass the distance $L_{source}$ not being splitted  is given by the exponential distribution,
\begin{equation}
\label{PSplitting}
P_{splitting}= \mbox{e}^{-\Gamma_{\gamma\to 3\gamma}\;\times \; L_{source}},
\end{equation}
Considering the initial photon flux instead of a single photon,  $P_{splitting}$ determines the suppression of the photon flux. 

\paragraph{Shower formation.}

In the subluminal LIV scenario any types of spontaneous photon decay are forbidden. Moreover, some photon decay processes in an  external field, allowed in the standard LI case, may become suppressed. The most important of these processes is the Bethe-Heitler process --- photon decay to an electron-positron pair in the Coulomb field of a nuclei. This process is 
crucial for the formation of atmospheric showers initiated by high-energy photons. 
The suppression of the Bethe-Heitler process in a subluminal LIV scenario was firstly estimated by \cite{Vankov:2002gt} and calculated in the model  in \cite{Rubtsov:2012kb}.  


As a result, photon-induced air showers in the subluminal scenario would start in the atmospheric deeper  than in the standard case. Very deep showers cannot be registered by the experiment. As the simplest criterium for the  registration we compare  the depth $X_0$ of the photon first interaction in the atmosphere with the total atmosphere depth $X_{\rm atm}$. If $ X_0 > X_{\rm atm}$, 
the mean shower starts near the ground surface, so the event will not be detected. Following this criterium, the probability for pair production of a photon in the atmosphere reads, 
\begin{equation}
\label{Preg}
P=\int_0^{X_{\rm atm}}dX_0
\;\frac{{\rm e}^{-X_0/\langle X_0\rangle_{LV}}}{\langle X_0\rangle_{LV}}\; = 1 - {\rm e}^{-X_{\rm atm}/\langle X_0\rangle_{LV}},
\end{equation} 
where the mean depth of the first interaction for the LIV case $\langle X_0\rangle_{LV}$ is expressed via the LI mean depth $\langle X_0\rangle_{LI}=57$ g $\mbox{cm}^{-2}$ and the ratio of the LI and the LIV Bethe-Heitler cross-sections. Taking into account the expression for LV Bethe-Heitler cross-section \cite{Rubtsov:2012kb}, one obtains 
\begin{equation}
\label{PShower}
P_{sh. sup.} = 1 -  {\rm exp}\left(- \frac{X_{atm}}{57\, \mbox{g}\, \mbox{cm}^{-2}}\left(\frac{7E_\gamma^4}{12m_e^2 M_{LV}^2}\right)^{-1}\log \frac{E_\gamma^4}{2m_e^2 M_{LV}^2}\right),  
\end{equation}
here the limit $ \frac{E_\gamma^4}{2m_e^2 M_{LV}^2} \gg 1$ is assumed.

\paragraph{Suppression of the observed photon flux.}
One can note that the different effects of photon splitting and suppression of shower formation appeared in different LIV scenarios (super- or subluminal) lead to the same result: the suppression of the observed photon flux. Thus, the prediction for observed flux in superluminal/subluminal LIV models reads,
\begin{equation}
\label{LVhyp}
\bigg(\frac{d\Phi}{dE}\bigg)_{LV}= P(E,M_{LV}) \times \bigg(\frac{d\Phi}{dE}\bigg)_{LI}\;.
\end{equation}
Here the suppression factor $P(E,M_{LV})$ is given by eq. (\ref{PSplitting}) for photon splitting and by eq. (\ref{PShower}) for shower formation.  The hypothesis of a certain $P(E,M_{LV})$ can be tested with experimental data. Note that these tests require a certain model of photon flux $\left(\frac{d\Phi}{dE}\right)_{LI}$ from the source. In order to set the most robust constraints on $M_{LV}$, in the case of more than one source models we choose the model with the maximal predicted flux.

\paragraph{Pair Production on CMB}

Another process important for LIV probing is pair production by a high-energy photon on soft photon backgrounds $\gamma\gamma_b \to e^+e^-$. This process occurs above a certain threshold, which for the background photon reads $\epsilon_{th} = \frac{m_e^2}{E_\gamma}$ in the LI scenario. In the presence of LIV the threshold gets modified \cite{Galaverni:2007tq}, 
\begin{equation}
    \label{ppCMB}
    \epsilon_{th} = \frac{m_e^2}{E_\gamma} \, \mp \, \frac{1}{4} \frac{E_\gamma^3}{M_{LV}^2}.
\end{equation}
Here the sign ``-'' refers to superluminal propagation while the sign``+''  to subluminal. The cross-section of the process above the threshold does not significantly depend on LIV \cite{Rubtsov:2012kb}. The  important outcome of this process is the attenuation of high-energy (TeV energy range) flux of extragalactic photons due the interaction with extragalactic background lights (EBL) \cite{Biteau:2015xpa}.  In the subluminal LIV scenario the pair production threshold is shifted to lagrer energies of background photons (related to smaller EBL density), making the extragalactic photon spectrum less attenuated \cite{Kifune:1999ex, Stecker:2001vb}. The absence of this effect  constrains the subluminal LIV scenario \cite{Martinez-Huerta:2020cut}. Considering galactic gamma rays in the $1 - 100$ TeV energy  range, one can neglect the pair production effect due to small distances inside the galaxy compared to the mean free path of pair production on the infrared background.   

For the galactic photons of sub-PeV energies observed by Tibet-AS$\gamma$ and LHAASO the threshold of pair production decreases to the energy range of cosmic microwave background (CMB), which density exceeds the EBL density by more than two orders of magnitude. The mean free path for a $\sim\,1$ PeV photon related to the pair production on CMB is $\sim\,10$ kpc \cite{Gould:1967zza,Vernetto:2016alq}, which is comparable to galactic scales. The galactic gamma-ray flux is attenuated by the factor $P_{p.p.  CMB}={\rm e}^{-L_{source}/L_{mfp}(E_\gamma)}$. In the subluminal LIV scenario the threshold increases above the energy of the CMB peak, and the attenuation disappears. Thus, in the subluminal LIV scenario the modification of pair production increases the observed gamma-ray flux while the shower suppression effect leads to the suppression. However, for the spectra observed in \cite{LHAASO,Amenomori:2021gmk}  the factor $P_{p.p.  CMB}$ cannot be less than $0.5$, so the effect of the modified pair production on CMB seems to be subleading compared to the shower suppression effect.

In the next two paragraphs we will consider the constraints from the observational data of Tibet-AS$\gamma$ \cite{Amenomori:2021gmk} and LHAASO  \cite{LHAASO}.

\paragraph{Tibet-AS$\gamma$.}


Tibet-AS$\gamma$ is a hybrid experiment devoted to the observations of VHE photons and cosmic rays, which is located at the altitude $4300$ meters above sea level in Tibet, China. The whole setup consists of an array of surface air shower (AS) detectors, combined with an underground array of muon detectors. The presence of a muon detector array yields a very high degree of gamma/hadron discrimination. In the recent work \cite{Amenomori:2021gmk} the Tibet-AS$\gamma$ collaboration presented gamma ray events from the dataset of 719 days of observation from the whole Tibet-AS$\gamma$ field of view with energies larger than $100$ TeV, grouped into three energy bins. 

In the last energy bin centered at $534$ TeV, 
$16$ gamma-ray-like events from the galactic disk (angle $|b|<5\deg$ from the galactic plane) against $1.35$ background events (statistical significance $6.0\,\sigma$) have been observed. The maximal energy of the observed gamma-ray event is $957^{+166}_{-141}$ TeV. All of these events are not connected with any known TeV source, so they are assumed to be diffuse gamma rays. According to modern theoretical models \cite{Lipari:2018gzn,Guo:2018wyf, Gaggero:2015xza}, it is assumed that these diffuse gamma rays are one of the products of hadronic interaction of charged cosmic rays in the interstellar medium. The measured diffuse flux in the last energy bin exceeds the model prediction \cite{Lipari:2018gzn} by the factor of two.
However, the authors of a recent work \cite{Koldobskiy:2021cxt} suggested the model of diffuse gamma rays from the outer Galactic disk which is in an agreement with Tibet-AS$\gamma$ data as well as with Fermi/LAT data on GeV-range diffuse gamma rays. 
In both superluminal and subluminal LIV scenarious the observed gamma-ray flux is suppressed compared to the model \cite{Koldobskiy:2021cxt}. We test the hypothesis of this suppression with the Tibet-AS$\gamma$ observational data.
For the calculation we have to specify a number of parameters. 
The maximal atmospheric depth at the Tibet location is $X_{atm} = 780$ g $\mbox{cm}^{-2}$ since the maximal zenith angle for Tibet-AS$\gamma$ events is $40$ degrees. Although the exact distance to the source (outer Galactic disk) for each diffuse gamma-ray event is unknown, one can estimate the distance with the maximal/minimum level in order to make the most conservative LIV constraint for a given process. Thus, considering the splitting process we take the minimal distance $L_{source} \approx 1$ kpc while for the pair production on CMB we take $L_{source} \approx 5$ kpc. The absorption coefficient related to the pair production on CMB for the last energy bin reads $P_{p.p. CMB} = 0.73$  \cite{Vernetto:2016alq}.
\begin{figure}[h]
    \centering
    \includegraphics[width=0.8\textwidth]{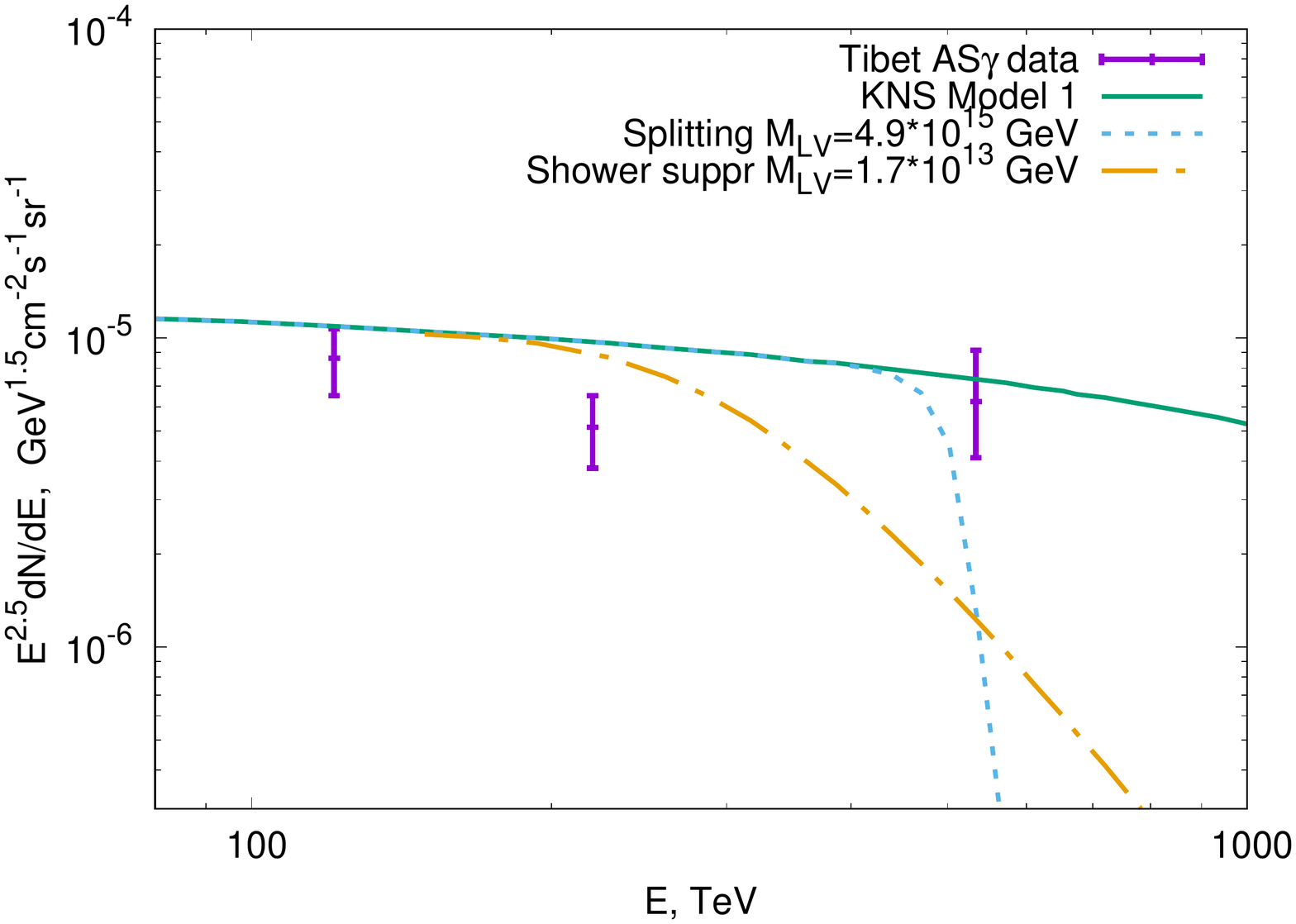}
    \caption{Diffuse gamma ray flux measured by Tibet-AS$\gamma$ from the outer Galactic disk (data points) \cite{Amenomori:2021gmk}, Model 1 of Koldobskiy et.al. \cite{Koldobskiy:2021cxt} for the flux from the outer Galactic disk (green solid curve), the prediction of superluminal LIV model with $M_{LV} = 6.6 \cdot 10^{15}$ GeV (blue dashed curve), the prediction of subluminal LIV model with $M_{LV} = 2.1 \cdot 10^{13}$ GeV (yellow dash-dotted curve).} 
    \label{fig:my_label}
\end{figure}

Diffuse gamma ray spectra of the outer Galactic disk measured by the Tibet-AS$\gamma$ collaboration, the theoretical model of the spectrum by Koldobskiy et.al. \cite{Koldobskiy:2021cxt} as well as the predictions of both LIV hypotheses   are presented at Fig. \ref{fig:my_label}.
First, one can notice that the cut-off of the spectrum in the splitting scenario is very sharp. 
On the other side, the spectrum steepening associated with the formation of deeper air showers is more smooth. However, even for the mean energy for the pre-ultimate bin the difference between LIV and LI scenarios is negligible. So, we test the hypothesis of a certain LIV mass scale only with the last bin with the $\chi^2$ test. The results are the following. The constraint associated with the absence of the splitting process (superluminal LIV scenario) reads  
\begin{equation}
\label{TibetSpli}
\mbox{(superluminal)}\quad M_{LV} > 4.9 \times 10^{15}\; \mbox{GeV},  \qquad 95\%\; \mbox{CL},
\end{equation}
while the constraint on the subluminal scenario (suppression of shower formation and pair production on CMB) reads
,
\begin{equation}
\label{BoundBetheHeitlerTibet}
\mbox{(subluminal)}\qquad M_{LV} > 
1.7\times 10^{13}\; \mbox{GeV}, \qquad 95\%\; \mbox{CL}.
\end{equation}
We set the photon decay bound with the lowest energy estimation for the highest energy photon event, $E=705$ TeV: at $95\%$ CL the energy of the event is larger than $E$. Substituting the energy to (\ref{PhDecay}), one obtains
\begin{equation}
\label{TibetDecay}
\mbox{(superluminal)}\quad M_{LV} > 4.9 \times 10^{14}\; \mbox{GeV},  \qquad 95\%\; \mbox{CL}.
\end{equation}
Note that the photon decay bound (\ref{TibetDecay}) is one order of magnitude weaker than the splitting one (\ref{TibetSpli}).

A hadronic origin of the diffuse gamma rays requires that the direct process of diffuse photon production is neutral pion decay, $\pi^0 \to \gamma\gamma$. In the scenario of superluminal LIV photons the pion decay process becomes kinematically forbidden if $M_{LV}>\frac{2E_\gamma^2}{m_\pi}$, where $m_\pi$ is the neutral pion mass (LIV in pions is assumed to be negligible). However, this bound is weaker than the photon decay bound (\ref{PhDecay}).

\paragraph{LHAASO} The Large High Altitude Air Shower Observatory (LHAASO) is a new-generation experiment designed for observation of photons and cosmic rays of very high energies located at the altitude 4410 meters in China. The whole installation consists of three components: a Kilometer Square Array (KM2A) of AS detectors (which include surface and underground muon detectors) connected with the Water Cherenkov Detector Array (WCDA) and Wide Field-of-view Cherenkov Telescope Array (WFCTA).  The observational data from less than a year of KM2A operation has been presented in a recent paper \cite{LHAASO}. The LHAASO collaboration declared \cite{LHAASO}  the observation of $12$ galactic gamma-ray sources above $100$ TeV energy. Particularly, the spectra of three gamma-ray sources, J2226+6057, J1908+0621 and J1825-1326, which continue up to $500$ TeV, have been provided in \cite{LHAASO}. In this paragraph we analyze this data and set the two-sided constraints on the mass scale $M_{LV}$. The photon decay and photon splitting constraints (both related to the superluminal scenario) from LHAASO data have been already obtained in \cite{Chen:2021hen} by a slightly different method; the results will be compared in this paragraph. We complement these constraints with shower formation constraints. The maximal zenith angle of LHAASO events is $50 \deg$ \cite{LHAASO}, so the maximal atmosphere depth for inclined showers at the LHAASO location is $X_{atm}\, =\, 890 \,\mbox{g}\,\mbox{cm}^{-2}$. 


It is shown in \cite{LHAASO} that  three observed source spectra have a  slight steepening above $100$ TeV compared to a power-law fit for the spectra, so the best fit for each of these spectra is a log-parabola curve. However, a power-law dependence is still not excluded. In order to estimate the maximal predicted flux at high energies we take the power-law fit. Based on this prediction we test both LIV hypotheses with experimental data (\ref{LVhyp}). 

\begin{figure}[h]
    \centering
    \includegraphics[width=0.49\textwidth]{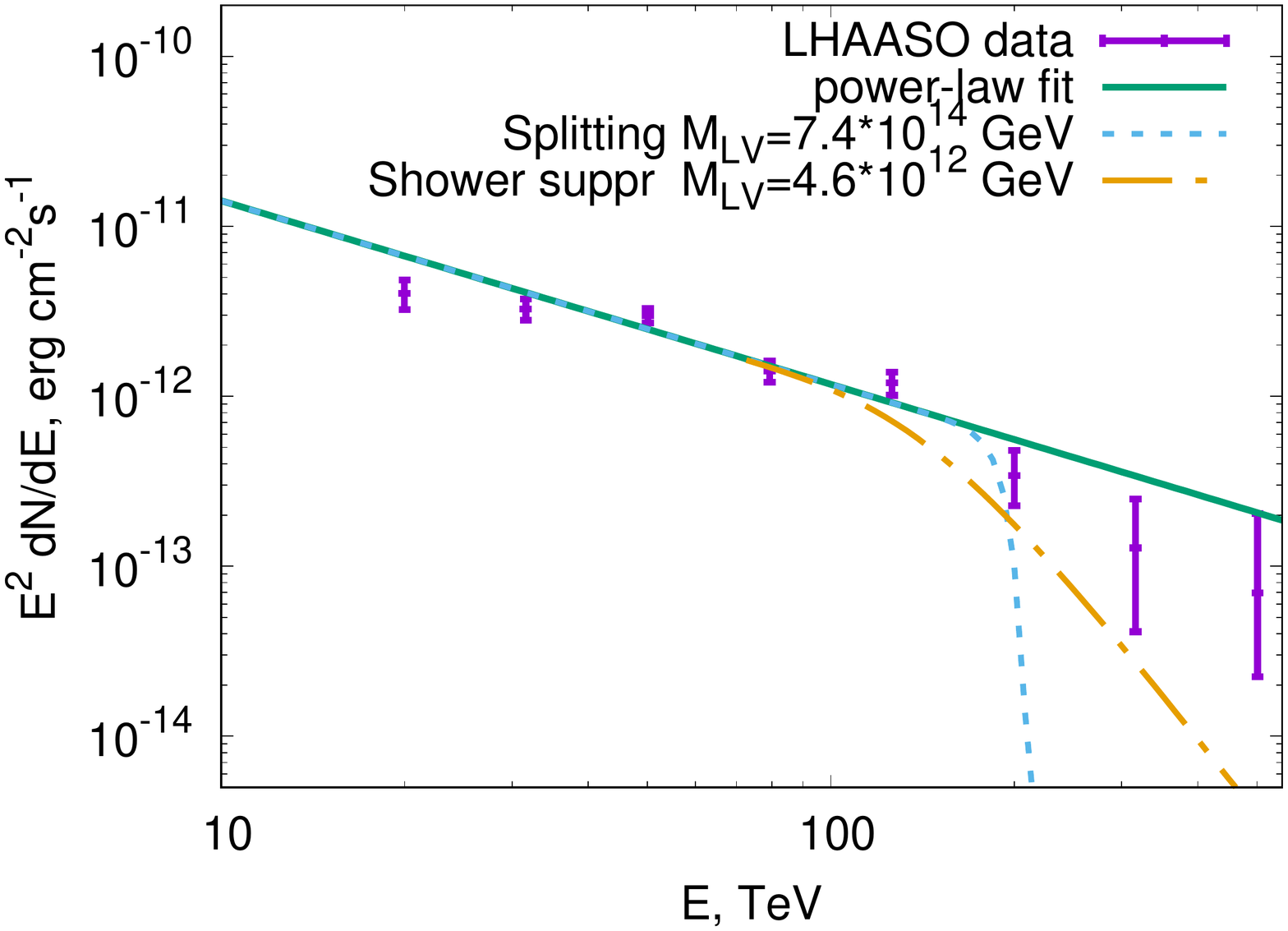} 
    \includegraphics[width=0.49\textwidth]{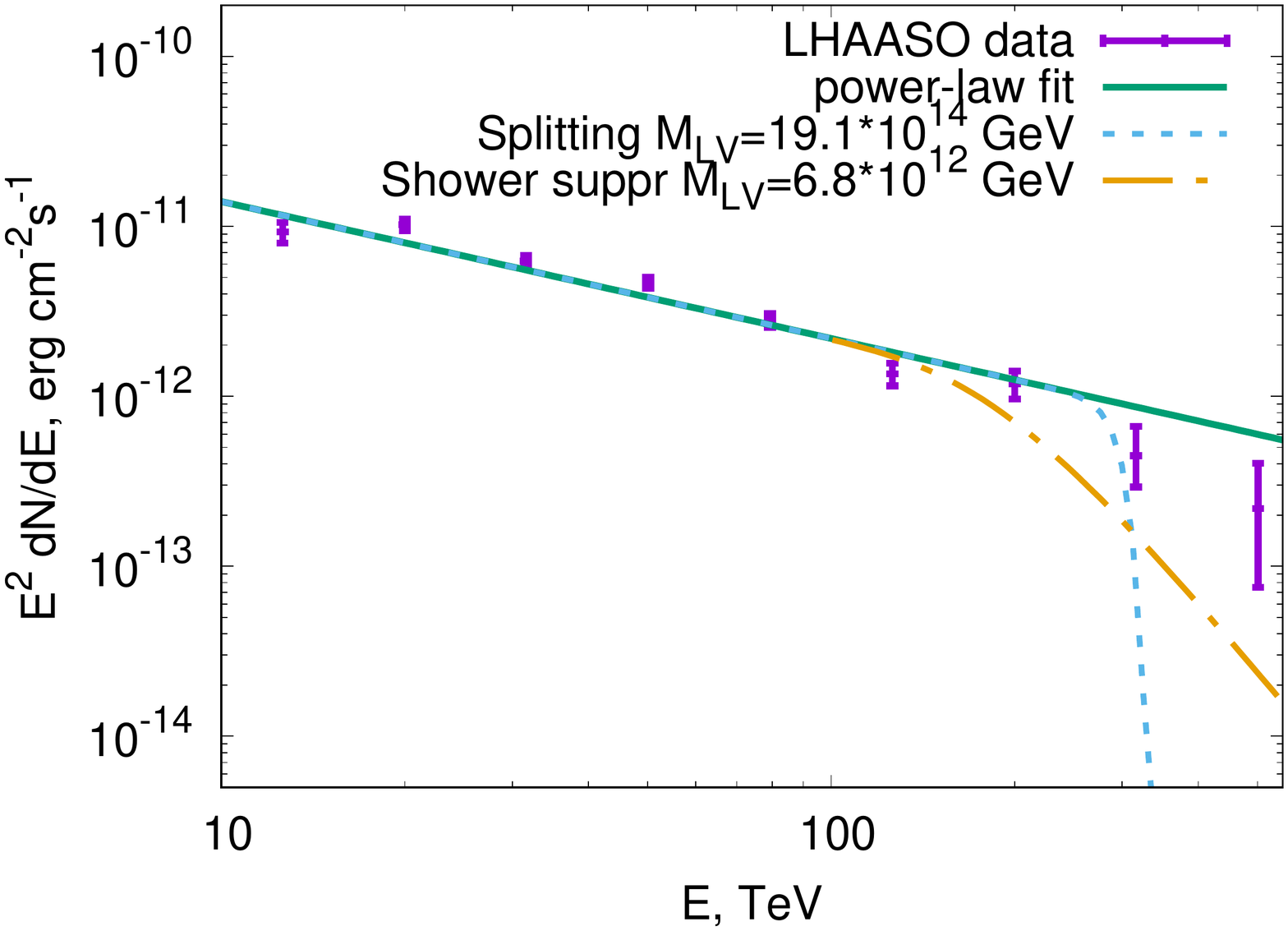} \\
    \includegraphics[width=0.5\textwidth]{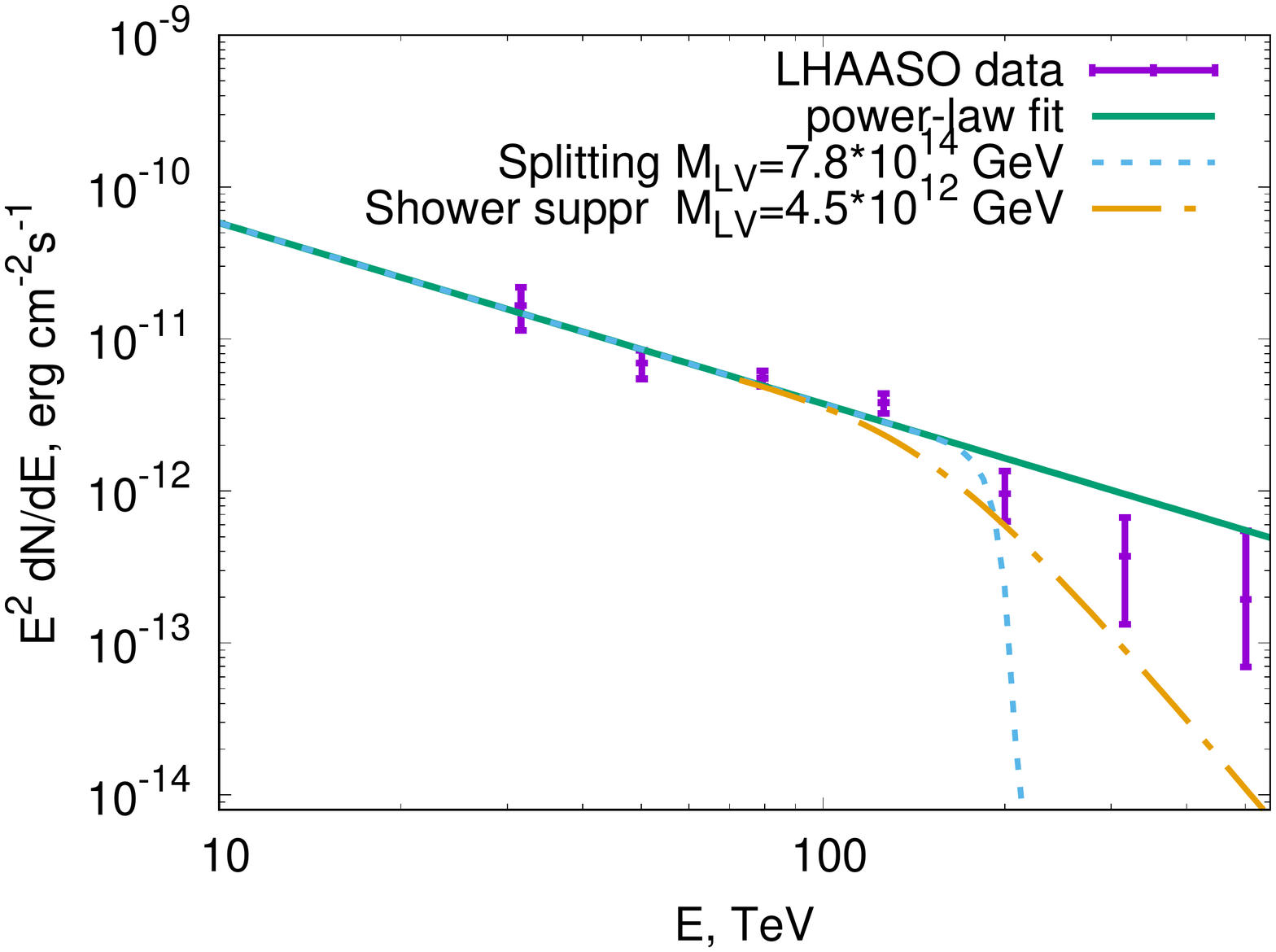} 
    \caption{The galactic source spectra measured by LHAASO (data points), the power-law fit for the spectra (green solid curve), the predicted flux under superluminal (blue dashed curve) and subluminal (yellow dash-dotted curve) scenarios. Top-left panel: J2226+6057, Top-right panel: J1908+0621, Bottom panel: J1825-1326. } 
    \label{fig:LHAASO}
\end{figure}

At the Fig. 2 we present differential energy spectra for each LHAASO source: experimental data points \cite{LHAASOdata}, the best power-law fit, and the prediction for the flux under the superluminal and subluminal LIV hypotheses. Note that the energy bins of LHAASO data are smaller than Tibet-AS$\gamma$ ones. The bin selection criterion for the LIV hypotheses test is as follows. On one side, we do not consider energy bins for which the difference between LI and LIV models are negligible. On the other side, we omit bins which statistical significance is less than two sigma since our aim is a $95\%$ CL exclusion limit -- the last bin for the source J1908+0621 and the two last bins for J2226+6057 and J1825-1326. It turns out that in order to test the splitting-induced cut-off a single bin for each source is relevant while for the shower suppression effect\footnote{Pair production on CMB is a subleading effect but is still taken into account. The absorption coefficients $P_{p.p. CMB}$ for the considered three LHAASO sources are presented at Fig. 6 of \cite{LHAASO}. } one has to analyze two bins for each source. 
Following these prescriptions we perform a $\chi^2$ test and constrain the mass scale $M_{LV}$ at $95\%$ CL for both superluminal and subluminal LIV scenarios. The results  are presented in Table 1.
\begin{table}[h]
    \centering
  \begin{tabular}{|c|c|c|c|c|}
\hline 
Source & L, kpc & subluminal $M_{LV}$, $10^{12}$ GeV & \multicolumn{2}{c|}{superluminal $M_{LV}$, $10^{14}$ GeV } \\ \hline
$\ $&$\ $&this work&this work & Chen et.al. \cite{Chen:2021hen}\\
\hline 
J2226+6057 & 0.8 & $4.6$ & $7.4$ & $14.6$ \\ 
\hline 
J1908+0621 & 2.37 & $6.8$  & $19.1$& $27.6$ \\ 
\hline 
J1825-1326 & 1.57 & $4.5$ & $7.8$ & $6.0$ \\ 
\hline 
\end{tabular} 
    \caption{The $95\%$ CL constraints on LIV mass scale from 3 sub-PeV sources observed by LHAASO. }
    \label{tab:LHAASO}
\end{table}

The difference in the superluminal scenario between our constraints and the ones obtained by Chen et al. \cite{Chen:2021hen} is less than a factor of two. The possible explanation is that  we take the most conservative power-law fit for the spectra while Chen et al. \cite{Chen:2021hen} used a log-parabola fit.

Let us also provide here the photon decay bound based on the gamma ray bound with the maximal energy obtained by \cite{Chen:2021hen},
\begin{equation}
M_{LV} > 1.4 \cdot 10^{15}\,\mbox{GeV} , \qquad 95\%\; \mbox{CL}.    
\end{equation}
Here the $95\%\; \mbox{CL}$ lower energy bound $E_\gamma = 1.21$ PeV for the maximal energy event has been taken.  

\paragraph{Discussion}

We have obtained two-sided constraints on the LIV mass scale from Tibet-AS$\gamma$ and LHAASO observations of gamma rays in the sub-PeV energy range. The constraint on $M_{LV}$ for the subluminal LIV scenario, based on Tibet-AS$\gamma$ data, $M_{LV} > 1.7\,\cdot \, 10^{13}$ GeV, (eq. (\ref{BoundBetheHeitlerTibet})) is the strongest in the literature. The previous constraint of this type  \cite{Satunin:2019gsl}, based on Tibet-AS$\gamma$ observation of Crab Nebula spectrum \cite{Amenomori:2019rjd}, is improved by a factor of 20. The constraint on the same parameter, based on the process of pair production on extragalactic background lights, is $ M_{LV} > 2.4\,\cdot \, 10^{12}$ GeV \cite{Lang:2018yog} --- one order of magnitude weaker than (\ref{BoundBetheHeitlerTibet}). 

The shower formation constraint is based on the model of the given spectrum of the source. Although there is a variety of models for the diffuse gamma ray flux from the Galactic disk, the significantly larger flux than proposed in \cite{Koldobskiy:2021cxt}  at sub-PeV energies would contradict to the measurements of the diffuse gamma rays in the GeV energy range as well as to observations of sub-PeV astrophysical neutrinos. 

It is worth noting that the method 
for constraining the suppression of shower formation based on a decreasing flux in a given energy bin leads to conservative bounds. Better constraints can be obtained by analysing the shower parameters (basically, the maximal depth $X_{max}$) for each photon-induced shower. This method has been proposed in \cite{Rubtsov:2013wwa} for constraining LIV on photons in $10^{19}-10^{20}$ eV energy range, which have not been detected yet. Inverting this argument, one can propose a unique signal for searching for subluminal LIV in atmospheric showers: the LIV photon shower would start deeper ($X_{max}$ would be larger) than the LI one, but the rest of photon shower parameters such as shower width and number of muons would not change.

\paragraph{Acknowledgements} The author thanks Dmitry Kirpichnikov, Valery Rubakov, Grigory Rubtsov and Sergey Troitsky for helpful discussions. The author would like to acknowledge the COST Action CA18108  ``Quantum gravity phenomenology in the multi-messenger approach''. This work is supported in the framework of the State project ``Science'' by the Ministry of Science and Higher Education of the Russian Federation under the contract 075-15-2020-778.


\end{document}